\documentclass[conf]{new-aiaa}
\usepackage[utf8]{inputenc}

\usepackage{graphicx}
\usepackage{amsmath}
\usepackage[version=4]{mhchem}
\usepackage{siunitx}
\usepackage{longtable,tabularx}
\usepackage{footnote}
\usepackage{url}
\usepackage{color}

\newcommand{\nemo}{\textit{NEMoSys}}

\title{Smart Adaptive Mesh Refinement with \nemo}

\author{Akash Patel\footnote{Research Engineer II, AIAA Professional Member, Email: patelakash9596@gmail.com}}
\affil{Illinois Rocstar LLC, Champaign, Illinois, 61820}
\author{Masoud Safdari\footnote{Adjunct Research Professor, Aerospace Engineering Department, Email: msafdari@illinois.edu}}
\affil{University of Illinois, Champaign, Illinois, 61801}

\begin{document}

\maketitle

\begin{abstract}
Adaptive mesh refinement (AMR) offers a practical solution to reduce the computational cost and memory requirement of numerical simulations that use computational meshes. In this work, we introduce a novel smart methodology for adaptive mesh refinement. Smart adaptive refinement blends classical AMR with machine learning to address some of the known issues of the conventional approaches. We provide an algorithm for adaptive refinement. Subsequently, we introduce a modular object-oriented structure for our smart AMR algorithm. Then we present procedures used for the training of a smart AMR model. The study follows with a demonstration of preliminary numerical studies indicating the feasibility of performing adaptive mesh refinement on a few demonstrative problems selected from the CFD domain. Finally, we conclude with a few comments about future work.
\end{abstract}

\section{Nomenclature}

{\renewcommand\arraystretch{1.0}
\noindent\begin{longtable*}{@{}l @{\quad=\quad} l@{}}
$Re$ &    reynolds number \\
$lr$& learning rate for machine learning network \\
$b_{0}$ & Initial bias for machine learning network \\
$W_{class_n}$ & class weights for output layer in machine learning network \\
$\nabla{U}^{'}$ & non-dimensional Velocity Gradient \\
$Q^{'}$ & non-dimensional Q-Criteria \\
$\omega^{'}$ & non-dimensional Vorticity \\
$\Delta^{'}$ & non-dimensional Modified Delta \\
\end{longtable*}}

\section{Introduction}
Preliminary algorithms for adaptive mesh refinement (AMR) were originally developed for computational fluid dynamics applications \cite{bergerPhDDissertationAdaptive1982}, and now AMR is routinely used in many disciplines such as material modeling \cite{vazAspectsDuctileFracture2001}, combustion \cite{pomraningModelingTurbulentCombustion2014}, biophysics \cite{oliveiraSimulationsCardiacElectrophysiology2016a}, astrophysics \cite{bryanENZOADAPTIVEMESH2014}, climate modeling \cite{chenExtendingLegacyClimate2019} and many others.AMR offers a practical solution to reduce the computational cost and memory requirements of mesh-based numerical analyses by eliminating the need for uniform refinement. Several AMR algorithms are developed, sharing a remarkable level of similarity. Efforts roughly split into two major categories: i) application-specific codes; and ii) library development. The latter poses significant challenges, as it should provide the flexibility to treat a diverse set of applications in a generic and reusable framework.

Development of software and algorithms for efficient, physics-independent adaptive mesh refinement dates back to 1970s. Early attempts set the ground work for two dimensional AMR of elliptic problems \cite{careyMeshrefinementSchemeFinite1976}, treatment of mid-edge (hanging) nodes with algebraic constraints \cite{careyAnalysisFiniteElement1976}, and penalty methods  \cite{careyPenaltyMethodsInterelement1982}. Later developments include introduction of bi-section method \cite{mitchellPLTMGSoftwarePackage1995} and extensions to three dimensions  \cite{plaza3DRefinementDerefinement2000}. More recent works were focused on leveraging rapidly evolving parallel computing infrastructure for AMR \cite{kirkLibMeshLibraryParallel2006c}. Parallelism is usually achieved through domain decomposition and efficient data structures for MPI communication of the partitioned mesh data \cite{kirkLibMeshLibraryParallel2006c}.

Traditionally, local error indicators obtained from a computed solution furnish automated adaptation. Several error indicators are developed for specific refinement purposes. Heuristic criteria are commonly used to increase the flexibility in library development efforts irrespective of the underlying physics of a simulation problem \cite{kirkLibMeshLibraryParallel2006c}. Similarly, more rigorous and problem-specific recovery indicators \cite{zienkiewiczSimpleErrorEstimator1987} are also widely used, but they should be considered as physics-independent criteria. Other ad hoc methods include feature-based indicators. They are considered competitive in resolving a-priori known critical feature(s) of the simulation, e.g., local vorticity fields \cite{kasmaiFeaturebasedAdaptiveMesh2011}. Local indicators are also extracted from a-posteriori error estimates \cite{verfurthPosterioriErrorEstimation1994a} that are closely linked to the residuals of numerical operators and governing equations. Regardless, a-priori knowledge about the continuous problem is still required. Finally, more precise indicators are developed, taking to account global measures of error. Evaluation of such indicators may involve additional computational efforts of solving a related dual (adjoint) problem \cite{bangerthAdaptiveFiniteElement2003}.

There are two major deficiencies of existing AMR algorithms for the treatment of both steady-state and time-dependent (or otherwise evolving) problems: i) the need for a-priori knowledge about the behavior of the solution field and ii) over-reliance on practitioner's expertise for criteria selection. AMR libraries merely rely on the user to pick an accurate error indicator for the problem at hand. An improper selection may lead to an efficiency lower than simple uniform refinement \cite{bangerthAdaptiveFiniteElement2003}. Further, often a combination of indicators is needed to treat complex problems. Similarly, the order of importance of employed error indicators or required thresholds for refinement based on a given criterion may evolve during transient analyses. Finally, derived quantities should be computed from the solution fields, increasing conventional AMR's computational cost. 

The need for refinement can be triggered by criteria other than error estimates. For instance, detection and refinement of the mesh at the neighborhood of non-slip boundaries can yield a better resolution of the flowfield. Currently, boundary refinement is performed manually based on user experience. Strofer et al. \cite{strofer2018data} propose the use of convolutional neural networks for the identification of types of vortex and boundaries of domains using steady-state flow solutions as training reference. Their approach is to map the mesh domain onto a 2D Cartesian grid and predict the boundaries, as well as vortex type present on those boundaries using CNN.

In this work, we investigate the application of machine learning in AMR. We aim to address challenges associated with classical approaches, in particular, the treatment of problems with no optimal refinement strategy. In the most simple form, adaptive mesh refinement can be cast into a logistic regression problem: a classifier computes whether refinement is needed or not.  We investigate several popular machine learning models as the decision-making apparatus for smart AMR methodology. We first introduce the algorithm within the context of classical adaptive mesh refinement. We then introduce the object-oriented design as well as the overall structure of our smart AMR. An implementation of this work for robust, automated adaptive mesh refinement is provided in our open-source meshing software \nemo\ \cite{nateshNEMoSysPlatformAdaptive2018}. Our study continues with a few preliminary studies to demonstrate the feasibility and robustness of the smart AMR. We select all problems from the CFD domain. We then wrap up with a discussion on some of the open issues, limitations, and plans for future extensions.

\section{Smart AMR Algorithm}\label{sec:algorithm}

\begin{figure}[hbt!]
	\centering
	\includegraphics[width=.27\textwidth]{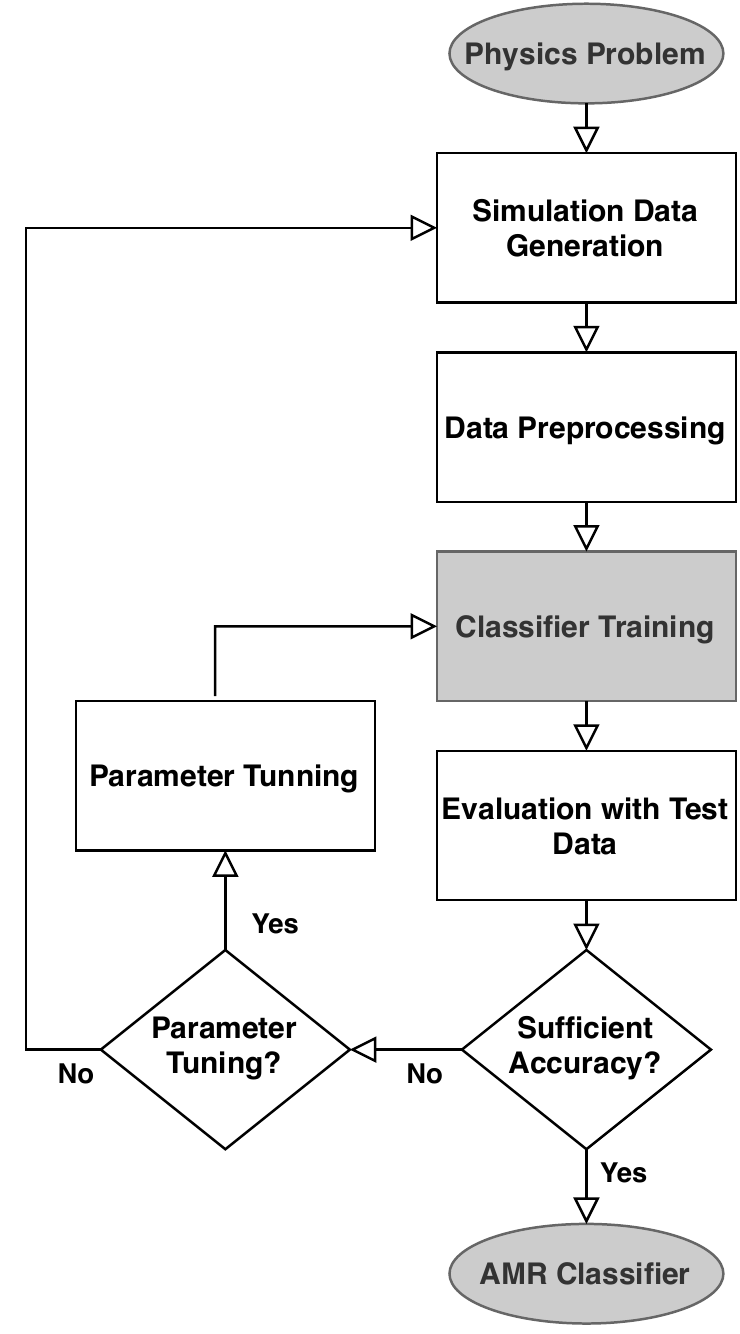}
	\caption{Smart refinement classifier training algorithm}
	\label{fig:alg}
\end{figure}

The algorithm we propose for smart AMR is summarized in Fig. \ref{fig:alg}. The algorithm involves three major steps: i) data generation, ii) model selection and iii) training and parameter tuning. Smart refinement is agnostic to the underlying PDEs and only requires properly selected training data. For steady-state problems, training data may include a reference solution to problems of similar nature to the desired one. Similarly, for a transient problem, a proper set of time snapshots of the solution field is sufficient. 

The following step is preprocessing. This step serves three purposes: i) data normalization, ii) computing error indicators, and iii) formating data for supervised training. The normalization process ensures the smoothness of the training procedure. We further explain our second step in Section \ref{sec:train}.

For simplicity, the refinement classifier operates on a point cloud. Regardless of the selected discretization scheme, a solution field can be easily represented by a point cloud, e.g. selecting a finite element solution's nodal values. In the current paper, we investigate structured meshes, and for simplicity, we include the point cloud's spatial coordinates as the input feature vector. A point cloud provides additional advantages by eliminating combinatorial irregularities and other complexities of meshes.


For each element, our smart AMR classifier marks (labels) the cell for refinement or coarsening, thus the input is solution field point cloud, and outputs are marks. We investigate the applicability of a trained classifier in problems of similar nature. In time-dependent problems, the classifier is trained on a few initial time snapshots. The possibility of such generalization is investigated in Sec. \ref{sec:app}.

\section{Implementation Aspects}
\begin{figure}[tbh]
	\centering
	\includegraphics[width=0.5\linewidth]{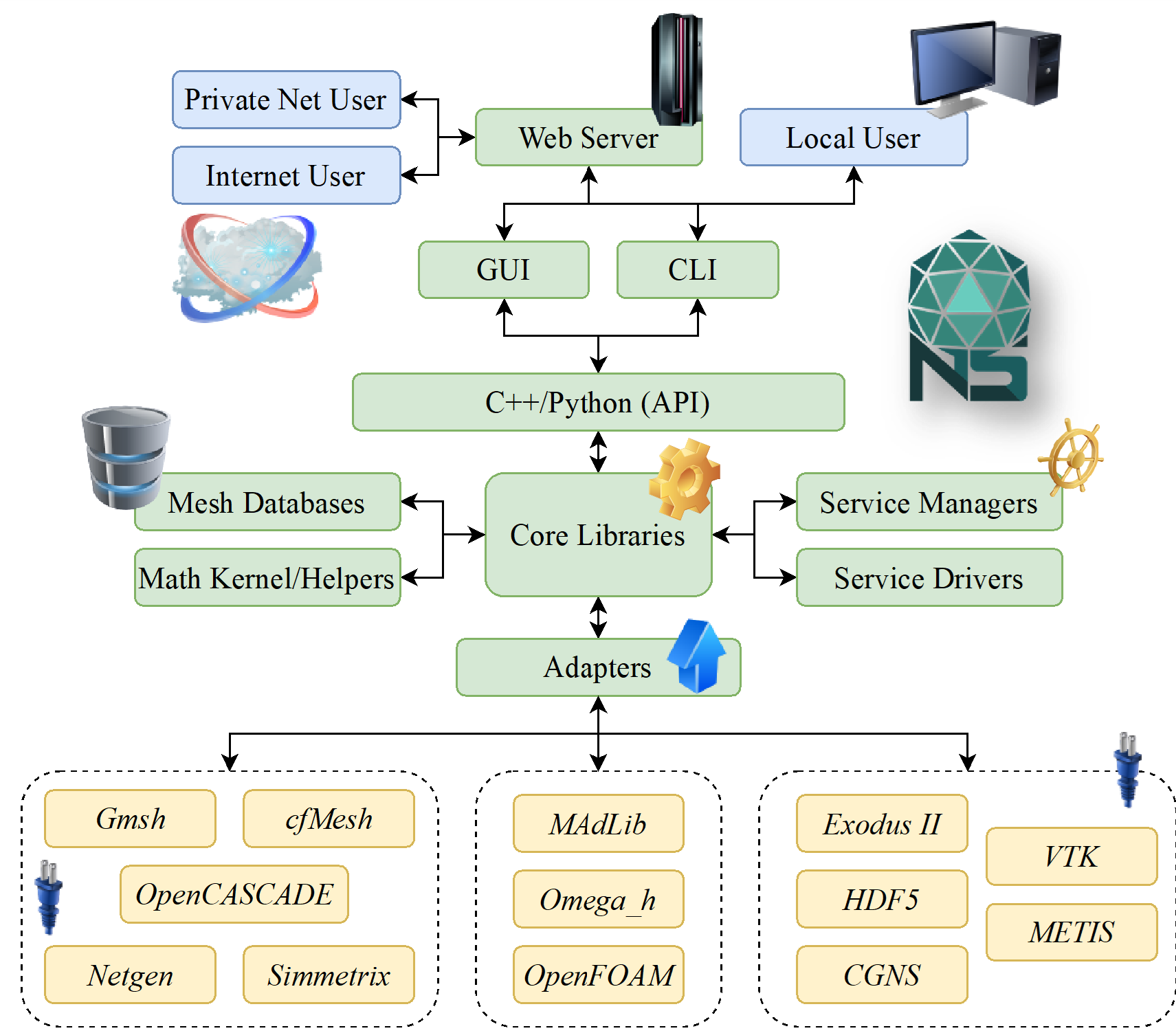}
	\caption{Modular architecture of \nemo\ with integrated libraries (bottom), core libraries/central API (center), and user end points (top)}
	\label{fig:nem_arch}
\end{figure}

\subsection{Object-oriented Design}\label{sec:design}
To facilitate ease-of-use, we implement our smart refinement as an extension to our open-source\footnote{\url{https://github.com/IllinoisRocstar/Nemosys}} meshing library \nemo. Figure \ref{fig:nem_arch} illustrates the modular and extensible architecture of the \nemo. The package is implemented in object-oriented C++ programming language. Several open-source meshing packages are integrated into the \nemo\ (shown at the bottom of Fig. \ref{fig:nem_arch}). We use \textit{OpenCASCADE}\footnote{\url{https://github.com/tpaviot/oce}} for I/O and B-Rep operations. \textit{Gmsh} \cite{geuzaineGmsh3DFinite2009a}, \textit{Netgen} \cite{schoberlNETGENAdvancingFront1997c}, and \textit{cfMesh} \footnote{\url{https://sourceforge.net/projects/cfmesh/}} serve mesh generation tasks. We use \textit{MAdLib} \cite{compereMeshAdaptationFramework2010} and \textit{Omega\_h} \cite{vittitowScalableDeterministicStateoftheArt}, and \textit{OpenFOAM} \cite{jasakOpenFOAMOpenSource2009} for h-adaptation. For mesh data storage and I/O, we use \textit{VTK} as the central data structure and provide support for mesh data I/O in \textit{Exodus II}, \textit{HDF}, and \textit{CGNS} formats. For local users, \nemo\ implements a  command-line interface (CLI). A network interface is also under development for remote accessibility. All interface modules communicate with the core library using a central python/C++ application programming interfaces (API) (shown at the center of  Fig. \ref{fig:nem_arch}).

\nemo\ is equipped with a pair of closely-interacting geometry (\textit{GModel}) and mesh (\textit{VtkDataSet}) databases. The \textit{GeoMeshBase} serves as the core object binding mesh and geometry data. Derivative objects (\textit{GmshGeoMesh}, \textit{VtkGeoMesh}, and \textit{ExoGeoMesh}) are concrete implementations of the \textit{GeoMeshBase} for additional mesh formats. In \nemo\ each service offered is controlled by a manager class that is a derivate of \textit{SrvBase} class. The services currently offered include i) mesh generation, ii) adaptive mesh refinement, iii) re-meshing, iv) quality control, v) format conversion, vi) solution data transfer, vii) solution verification, viii) simulation input generation, ix) partitioning, x) joining, and xi) nuclear mesh generation.

\textit{AMRFoam} is our preliminary implementation of the smart AMR procedure. This object provides a dynamic mesh data structure and implements methods for different steps of the refinement. \textit{AMRFoam} takes a trained AMR classifier prepared according to the procedure described in Section \ref{sec:train}. For each cell,  \textit{AMRFoam} evaluates the classifier to compute the probability of refinement/coarsening. The mesh adaptation is then performed by another method implemented by this class of delegated to the physics solver. The class also provides optional capabilities for solution data transfer and file I/O. For improved automation, \textit{AMRFoam} also takes information such as the starting and ending time steps for refinement, the maximum allowable number of elements, maximum permissible levels of refinement/coarsening, and I/O frequency.

\subsection{Feature Extraction}\label{sec:criteria}
Smart AMR is best trained to perform refinement based on a set of characteristic features of the solution field. In the current work, we use vorticities as characterisic feature for refinement. A vortex structure in a flowfield has no univocal mathematical definition. It can be described as the rotating motion of fluid around a common centerline. In contrast, we may be able to easily identify a vortex by visualizing a flowfield, however perception about where the vortex ends vary among analysts. To automate the adaptive mesh refinement process, identifying flow features without the visual means and \textit{a-priori} knowledge of flow solution is of utmost importance. Although a vortex is commonly perceived as a circular motion, shear and boundary layer vortex flows do not necessarily follow clear circular motions. Several criteria such as Q-criterion, $\lambda_2$ criteria, and $\Delta$ criteria have been developed to visualize such vortex structures. Vorticity ($\omega$) is also used to identify turbulent structures in a flow. These criteria work well for a given flow problem, but due to their representation of vortex structures in localized dimensioned values, their thresholds are not easily generalizable across multiple flow problems. As such, none of these criteria can be considered optimal for adaptive refinement.

\begin{figure}[htb]
	\centering
	\includegraphics[width=0.7\linewidth]{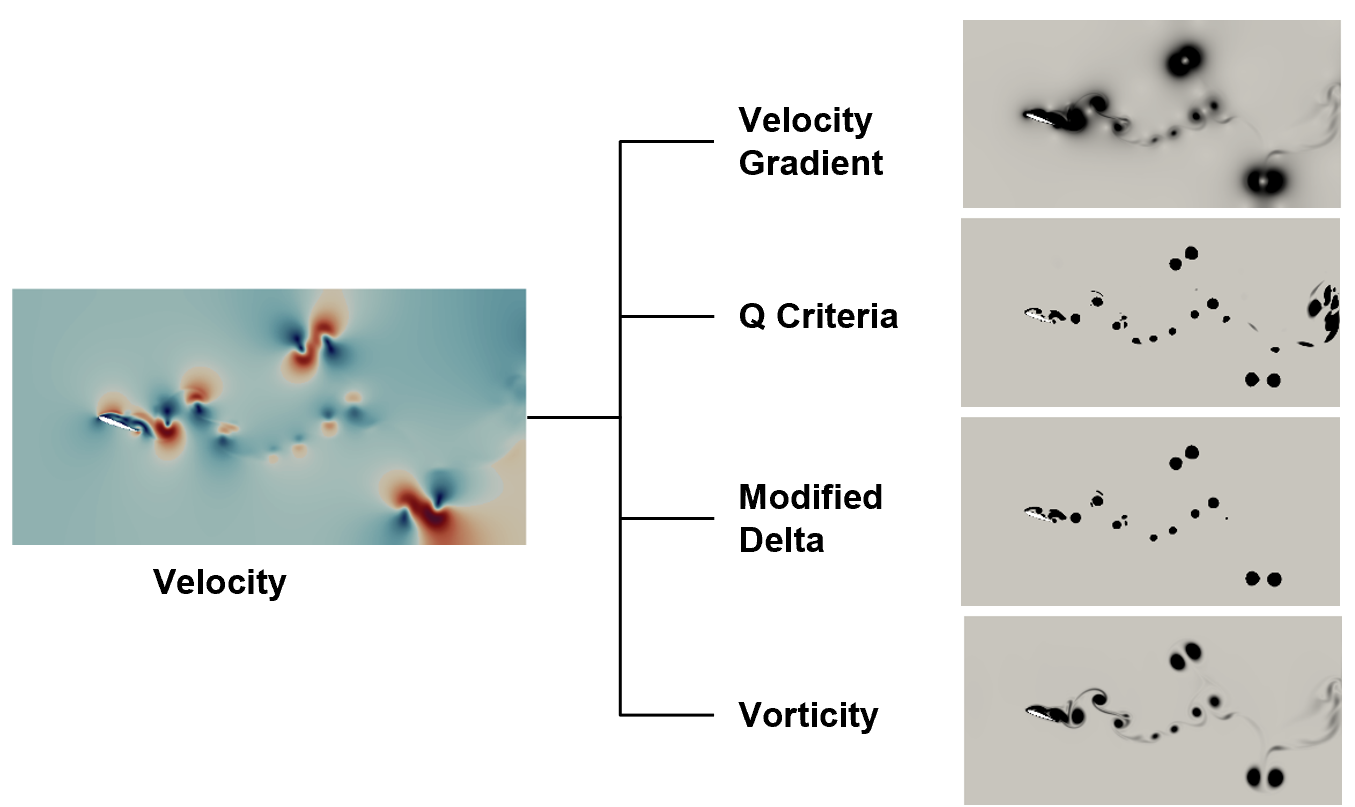}
	\caption{Visualization of non dimensional quantities selected for automatic feature extraction}
	\label{fig:ref_quant}
\end{figure}

Therefore, we look for non-dimensional criteria that capture turbulent flow structures across various flow problems without large variation in thresholds. Kamkar and Wissink \cite{kamkar2009automated} have proposed non-dimensional form of Q-criteria, and a modified, non-dimensional version of $\Delta$ criteria. These criteria use local velocity strain tensor and yield good results in capturing flow structures of importance with minimal variation in threshold values. Along with non-dimensional Q-criteria and modified $\Delta$, we propose normalized velocity gradient and vorticity. We normalize the velocity gradient and vorticity with their respective local infinite norms. We use these criteria to train our proposed machine learning model implemented in \textit{AMRFoam}. We believe that these quantities best represent several key features where refinement is most needed in a CFD problem. Figure \ref{fig:ref_quant} shows all selected quantities on a domain surrounding the airfoil at 20$\si{\degree}$ angle of attack.

\subsubsection{Velocity Gradient}\label{para:vGradient}
Velocity gradient represents localized change in velocity in all three flow directions. Velocity gradient values vary with flow velocity and cannot be used directly as general purpose refinement criteria. We normalize velocity gradient by its infinite norm in the entire domain. This is given by,
\begin{equation}\label{eqn:vGrad}
\nabla{U}^{'} = \frac{\|u_{i,j}\|}{\|{u_{i,j}}\|_{\infty}},
\end{equation}
where $u_{ii}$ are the components of the velocity gradient field. Normalized velocity gradient shows very narrow variation in threshold for detecting vortex structures across several different flow problems.
\subsubsection{Q-Criteria}
The Q-criteria is defined as a measurement of difference between velocity rotation and velocity strain magnitudes. The dimensional version of it depends on velocity value and characteristic length.
\begin{equation}
Q = \frac{1}{2} (||\Omega||^{2} - ||S||^{2})
\end{equation}
where, $\Omega$ is anti-symmetric part of velocity gradient and $S$ is symmetric part of velocity gradient. This is normalized by symmetric part of velocity gradient and normalized form of Q criteria is defined by,
\begin{equation}
Q^{'} = \frac{1}{2} \left(\frac{||\Omega||^{2}}{||S||^{2}} - 1\right)
\end{equation}
Non-dimensional Q criteria shows the quality of rotation. Larger area (\ref{fig:ref_quant}) shows presence of a stronger vortex and smaller areas indicate a weaker vortex.
\subsubsection{Modified $\Delta$}
Modified $\Delta$ is derived from eigenvalues of velocity gradient tensor as described in \cite{kamkar2009automated}. The solution of eigenvalue problem of velocity gradient ($\nabla\vec{U}$) for a 3D case is given by roots of equation, 
\begin{equation}\label{eq:deltaEqn}
\lambda^{3} + P\lambda^{2} + Q\lambda + R = 0
\end{equation}
where $P = -trace[\nabla\vec{U}]$, $Q = \frac{(P^{2} - trace[(\nabla\vec{U})^{2}])}{2}$, and $R = -det[\nabla\vec{U}]$. The solution to equation \ref{eq:deltaEqn} will either yield 3 real roots $(\lambda_1, \lambda_2, \lambda_3)$, or a real root with a pair of complex roots ($\lambda_1, \lambda_{cr} \pm \lambda_{ci}$).

Complex roots indicate that a local vortex motion exists and it's strength is indicated by imaginary part of complex root pair. This strength indicator is normalized by symmetric part of velocity gradient ($||S||$) and non-dimensional form of modified $\Delta$ criteria is given by,
\begin{equation}
\Delta^{'} = \frac{\lambda_{ci}}{||S||}
\end{equation}
\subsubsection{Vorticity}
Vorticity is described as a curl of velocity. Vorticity in its raw form helps visualize a swirling motion occuring in a flow field. Threshold for identifying such phenomena depends on local values of velocity and length scales. We normalize vorticity is by its infinite local norm to normalize it. The non-dimensional vorticity therefore, is calculated as,
\begin{equation}
\omega^{'} = \frac{\nabla \times \vec{U}}{||\nabla \times \vec{U}||_{\infty}}
\end{equation}

\subsection{Classifier Architecture and Training}\label{sec:train}
\nemo\ integrates with \textit{tensorflow}\footnote{\url{https://www.tensorflow.org/}} for low-level training and evaluation of smart AMR classifiers. In this study, we use two different classifiers, a fully connected artificial neural network (ANN) with four input neurons, and a convolutional neural network (CNN) with 4 channels of $5 \times 5$ pixel matrix.
\begin{figure}[htb]
	\centering
	\includegraphics[width=0.6\linewidth]{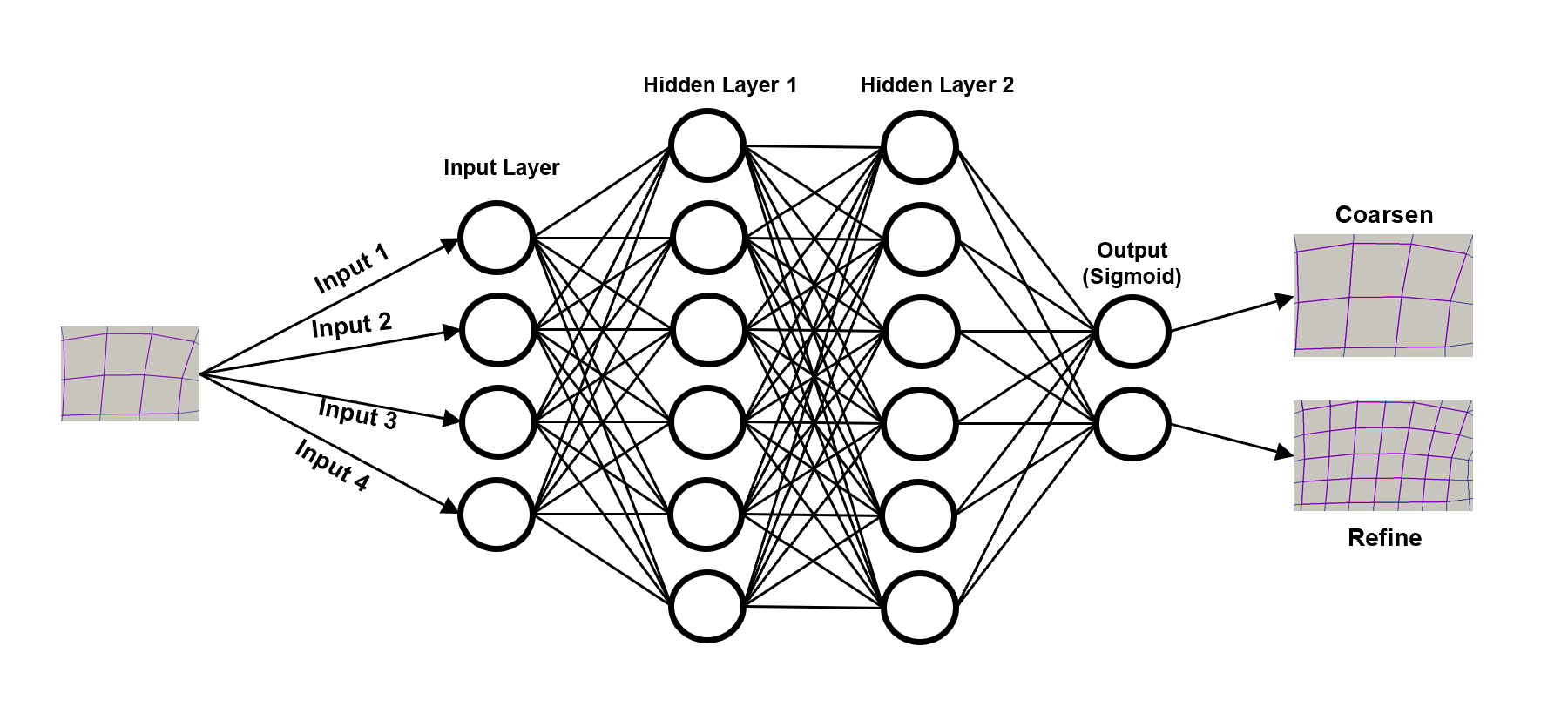}
	\caption{Architecture for artificial neural network (ANN) classifier}
	\label{fig:NN_Arch}
\end{figure}
For CFD problems, we consume two different sets of inputs for each cell. Normalized values of the coordinates of the centers and velocity gradient (\ref{eqn:vGrad}) or four different non-dimensional features representing quantities discussed in section \ref{sec:app}. A general architecture for the neural network is shown in Figure \ref{fig:NN_Arch}. For neural network, input layer is connected with $n$ fully connected hidden layers denoted by $X_1,\cdots, X_n$. We use one output neuron connected to the hidden layers for binary classification and n output neurons for classification between total $n\ (> 2)$ classes.

\begin{figure}[htb]
	\centering
	\includegraphics[width=0.8\linewidth]{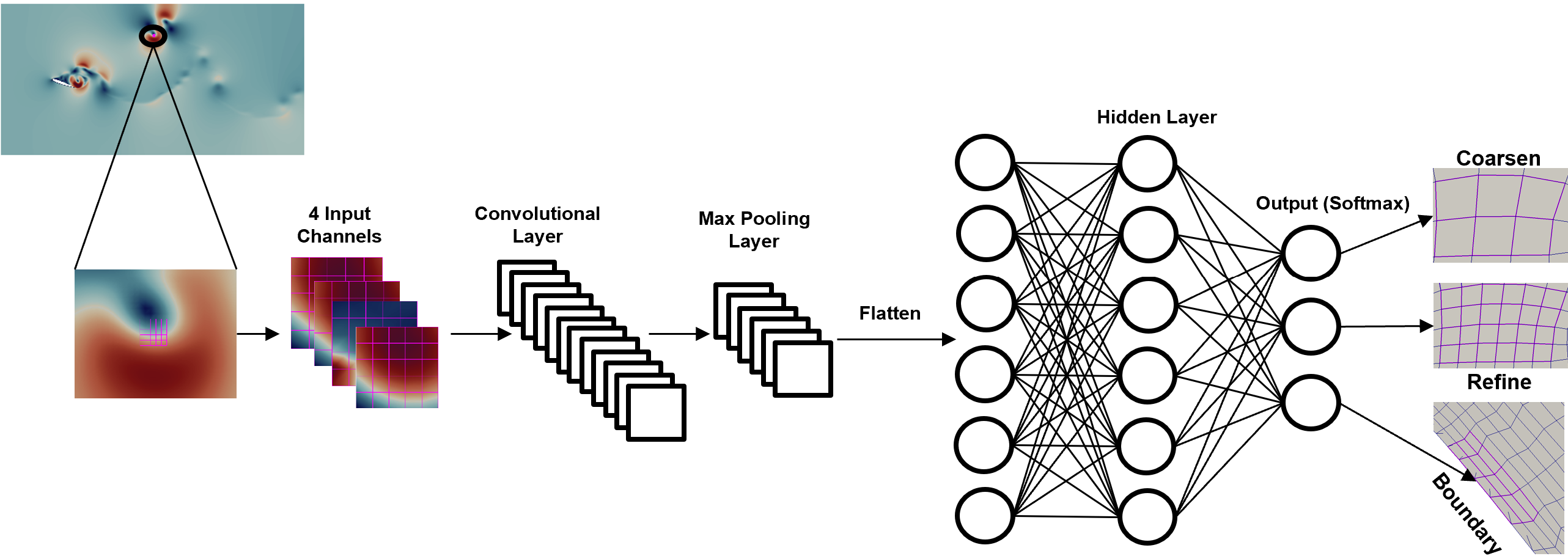}
	\caption{Architecture for convolutional neural network (CNN) classifier}
	\label{fig:CNN_Arch}
\end{figure}

For convolutional neural network, input layer consists of 4 channels of $5 \times 5$ pixel matrix. This matrix is comprised of a cell at center and its 24 neighbors (two rings) arranged in spatially correlated positions. Each channel represents one of four non-dimensional quantities. Input layer is connected with 2 layers of convolution and max pooling operations followed by one hidden layer connected to output neurons as shown in Figure \ref{fig:CNN_Arch}. Output neuron computes the probability of refinement. During model construction and training, we found that for problems studied, the combination of ReLu $(f(x) = max(0,x))$ activation function for hidden layers and Sigmoid $(f(x) = \frac{1}{1+e^{-x}})$ activation function for output layer yields best performance for binary classification problems. For multi-class prediction, we use Softmax $(f(x) = \frac{e_{x_i}}{\sum_{j=1}^{K} e^{x_j}})$ activation function for output layer.

In an adaptive refinement, the number of refinement and coarsening locations rarely balance; coarsening labels usually outnumber refinement labels by large. The lack of balance warrants some adjustments in the architecture of the classifier to avoid potential overfitting problems. To achieve balance, we use an initial bias and class weights. We calculate initial bias $(b_0)$ using,
\begin{equation}
b_0 = log_e \left(\frac{r}{c}\right),
\end{equation}
\noindent
where $r$ is the number of refinement labels, and $c$ is the number of coarsening labels. 

We use initial bias for binary classification problems only. For multi-class classifications, only class weights are used. Class weights are means of telling machine learning networks to focus on under-represented classes more during training so that model does not overfit on over-represented classes. For class weights in the output layer we use,
\begin{equation}
W_{class_n} = \frac{1}{l_n} \left(\frac{l_1 + l_2 + ... + l_n}{n}\right)
\end{equation}
\noindent
where $l_n$ is number of labels for class n.
The training of our neural network involves finding a minima of the loss function using a stochastic gradient descent procedure. To ensure that model does not overshoot the global minimum, learning rate (lr) is controlled using time inverse decay given by
\begin{equation}\label{eqn:lr_schedule}
lr = \frac{l_0}{1 + d_r \left(\frac{g_{step}}{d_{step}}\right)},
\end{equation}
\noindent
where $l_0$ is the initial learning rate, $d_r$ is the decay rate, $g_{step}$ is the global step (number of training points used in one batch), and $d_{step}$ is the decay step (total number of batches). Once the training is successfully completed, the trained model is used by \textit{AMRFoam} class for smart refinement, as discussed in \ref{sec:design}.

\section{Preliminary Feasibility Studies}\label{sec:app}
\subsection{Pitz-Daily Problem}\label{sec:pitz}
\begin{figure}[htb]
	\centering
	\includegraphics[width=0.5\linewidth]{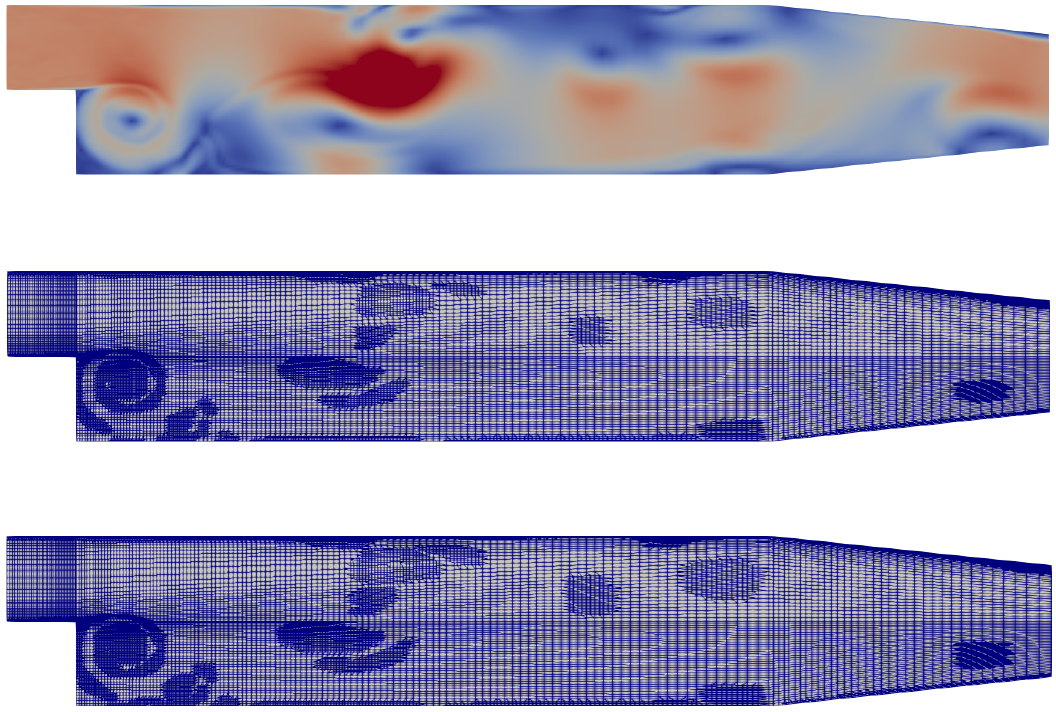}
	\caption{Flow field velocity for the canonical Pitz-Daily problem (Top). Expected refined mesh for capturing vorticities (Center). On-the-fly smart adaptive refinement for the same time step calculated with 98\% accuracy (Bottom)}
	\label{fig:ml_refinement}
\end{figure}

For the first demonstration case, we apply smart AMR to the canonical Pitz-Daily problem.  Figure \ref{fig:ml_refinement}-(Top) shows the velocity field computed at a time step dominated by vortices formed at the wake of the step. The problem was set up and ran with \textit{pimpleFoam} solver from \textit{OpenFOAM}. The CFD solver was augmented to use \textit{AMRFoam} class.  The Reynolds number for the flow field was selected to be $20,000$. Problem runtime was selected to be 0.1 seconds with a time step of $10^{-3}$ second. We opt for vorticity as the dominant feature of this physics simulation using insight obtained from the velocity gradient and vorticity field for the first 100 time steps. Thus as the input feature vector, we use coordinates scaled by the extents of the domain and the magnitude of the velocity gradient tensor scaled by its infinite norm in the entire domain given by \ref{eqn:vGrad}.
\noindent

\noindent
An artificial neural network with four input neurons, two fully connected hidden layers with fifteen and ten neurons, and one output neuron was used as the classifier. The classifier was trained on data collected from the time step range 1-25 (up to $0.025$ second of runtime). The neural network training was performed on data obtained from a higher density mesh (\textasciitilde79K cells) compared to the coarser mesh (\textasciitilde11.5K cells) shown in Fig. \ref{fig:ml_refinement} on which the predictions were made. We used 30\% of training data for cross-validation of the classifier after each epoch of training. We performed training until the loss function stops improving for ten consecutive epochs.

\begin{figure}
	\centering
	\includegraphics[width=0.7\linewidth]{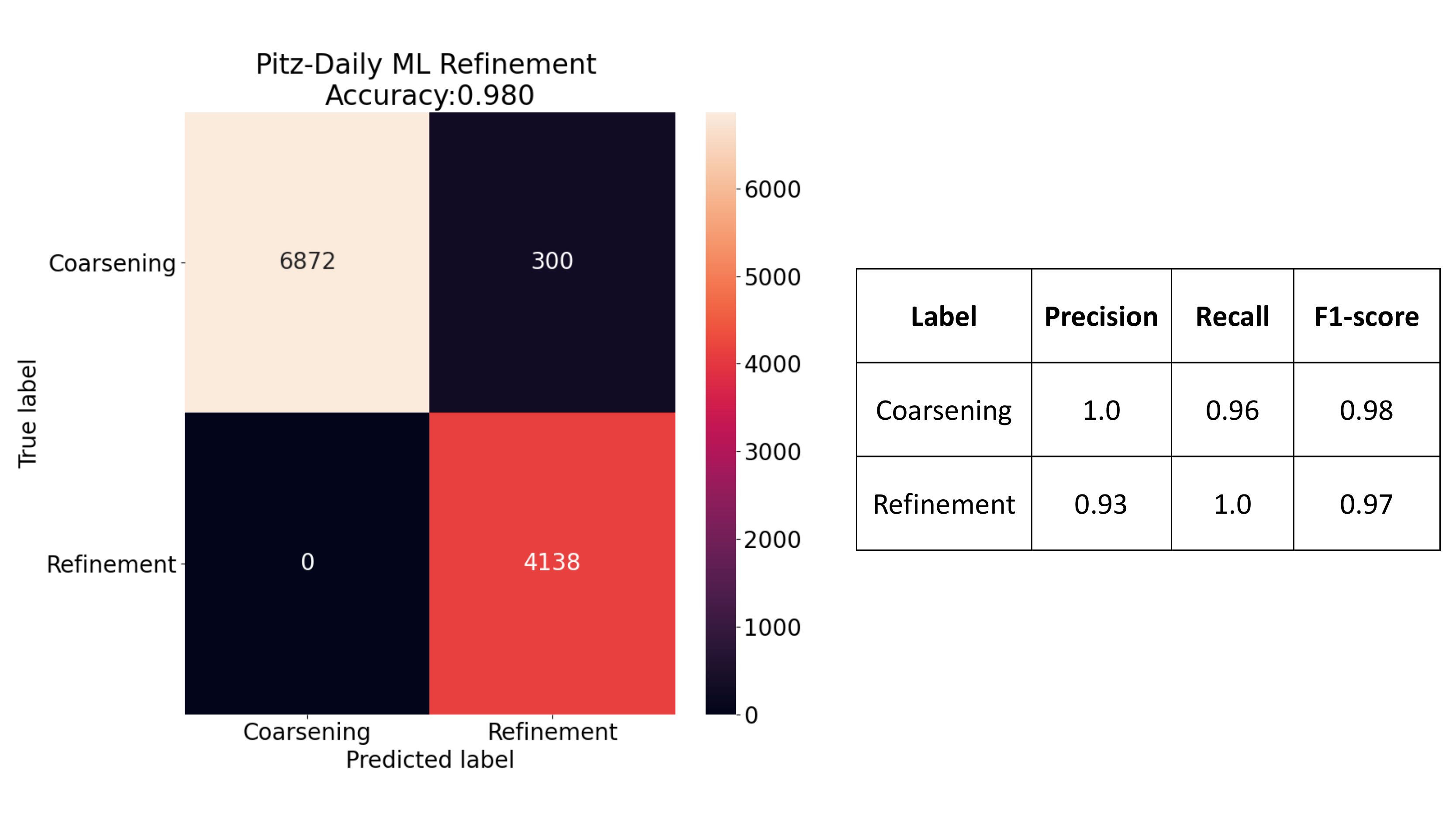}
	\caption{Confusion matrix for Pitz-daily case (Left), and precision, recall, and f1-scores (Right)}
	\label{fig:pitz_cm}
\end{figure}

Figure \ref{fig:ml_refinement}-(Center) shows expected refinement for $t=0.05$ second computed using a conventional value-based refinement procedure. The mesh was refined for best resolving vorticities. Figure \ref{fig:ml_refinement}-(Bottom) shows the smart refinement prediction for the same time step. The smart refinement was able to achieve 98\% accuracy in predictions compared to conventional value-based refinement. Fig \ref{fig:pitz_cm}-(Left) shows the confusion matrix for the prediction indicating that the classifier mispredicted 300 coarsening labels out of a total of 7172 labels and properly predicted all refinement labels. It should be noted that mispredicted coarsening labels slightly reduce the AMR efficiency, whereas the accuracy of the physics simulation is preserved. Figure \ref{fig:pitz_cm}-(Right) lists the precision, recall, and f1-scores for the predictions generated by the classifier.

\subsection{Turbofan Engine Blades}\label{sec:fan}
\begin{figure}[htb]
	\centering
	\includegraphics[width=0.5\linewidth]{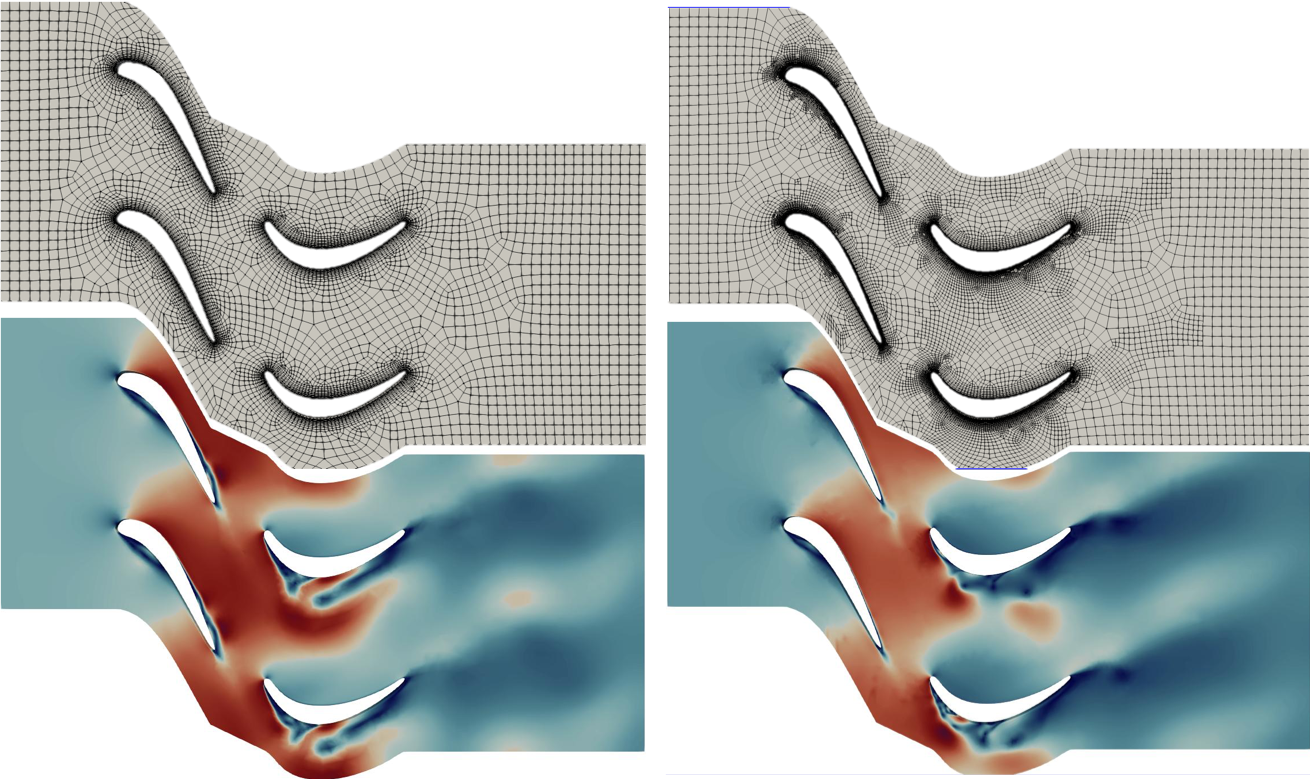}
	\caption{Flow field captured by an unrefined mesh (Left), and a mesh refined by smart refinement methodology (Right). Vorticities are better resolved by the refined mesh}
	\label{fig:smart_amr_blade}
\end{figure}

As discussed in Section \ref{sec:algorithm}, a trained classifier can be applied to an ensemble of simulation problems with similar physical characteristics. We investigated the feasibility of such applications through an example. Figure \ref{fig:smart_amr_blade}-(Left) illustrates the initial mesh and flow field computed for a turbofan engine with a pair of fan blades and guiding vanes an inlet at left, an outlet at the right, and periodic boundary conditions at the top and the bottom. Reynolds number of the flowfield was selected to be 600,000, and the problem was solved for a total runtime of $0.002$ second. We employed the same ANN classifier described in Section \ref{sec:pitz}. In these problems, vorticities were predominantly observed in the wake of blades and vanes, so a feature-based AMR with proper criteria for vortex identification such $Q$, $\omega$, $\delta$, and $\lambda_2$ criteria can be employed. Similarly, the magnitude of the velocity gradient tensor itself can be considered as a proper refinement criterion. 

\subsubsection{Pure Classifier}
The classifier used in this problem is similar to the one used in Pitz-Daily problem. This classifier was trained using simulation data collected from a much simpler airfoil flow field simulation shown in Fig. \ref{fig:ref_quant}. The training problem includes an airfoil at a 20\si{\degree} angle of attack with the Reynolds number of $100,000$. As observed, the training problem shares similarities with the primal problem, but it is much simpler to set up and run than the primal problem. The neural network was trained with a total of $\sim$11.1M inputs (mesh cell center coordinates and non-dimensional velocity gradient) collected from the time-evolving airfoil problem. As shown in Fig \ref{fig:smart_amr_blade}-(Right), the smart refinement methodology has was able to identify and resolve the vortex zones successfully.

\subsubsection{Ensemble Classifier}
\begin{figure}[htb]
	\centering
	\includegraphics[width=1.0\linewidth]{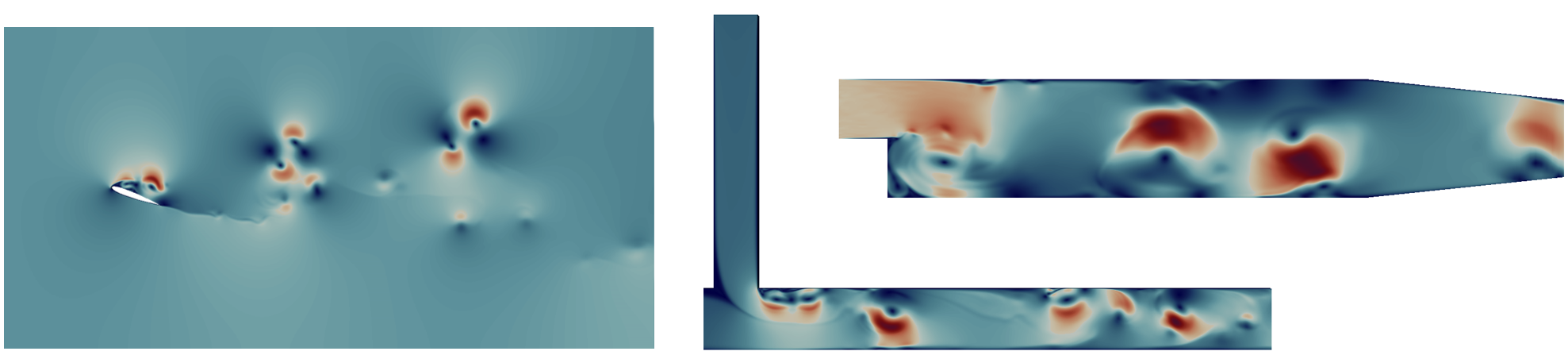}
	\caption{Flowfield for the flow over airfoil at 20\si{\degree} AOA (Left); Flowfield for T-junction problem (Center); Flowfield for Pitz-daily problem (Right)}
	\label{fig:allproblems}
\end{figure}

To assess the degree of generalization this neural network can offer across different flow problems, we added Pitz-Daily and T-Junction (Figure \ref{fig:allproblems}) flow solutions along with flow over airfoil solution to the current neural network training dataset. Training the classifier on this new dataset, we observed a drop in training accuracy from 98\% to 89\%. Upon further exploration, we identified input coordinates as a root cause of the issue. We then replaced mesh cell center coordinates with three non-dimensional quantities (Q-criteria, Vorticity, and Modified $\Delta$) described in \ref{sec:criteria}. Now the input layer consists of non-dimensional velocity gradient, vorticity, modified $\Delta$, and Q criteria. 

\begin{figure}[htb]
	\centering
	\includegraphics[width=0.8\linewidth]{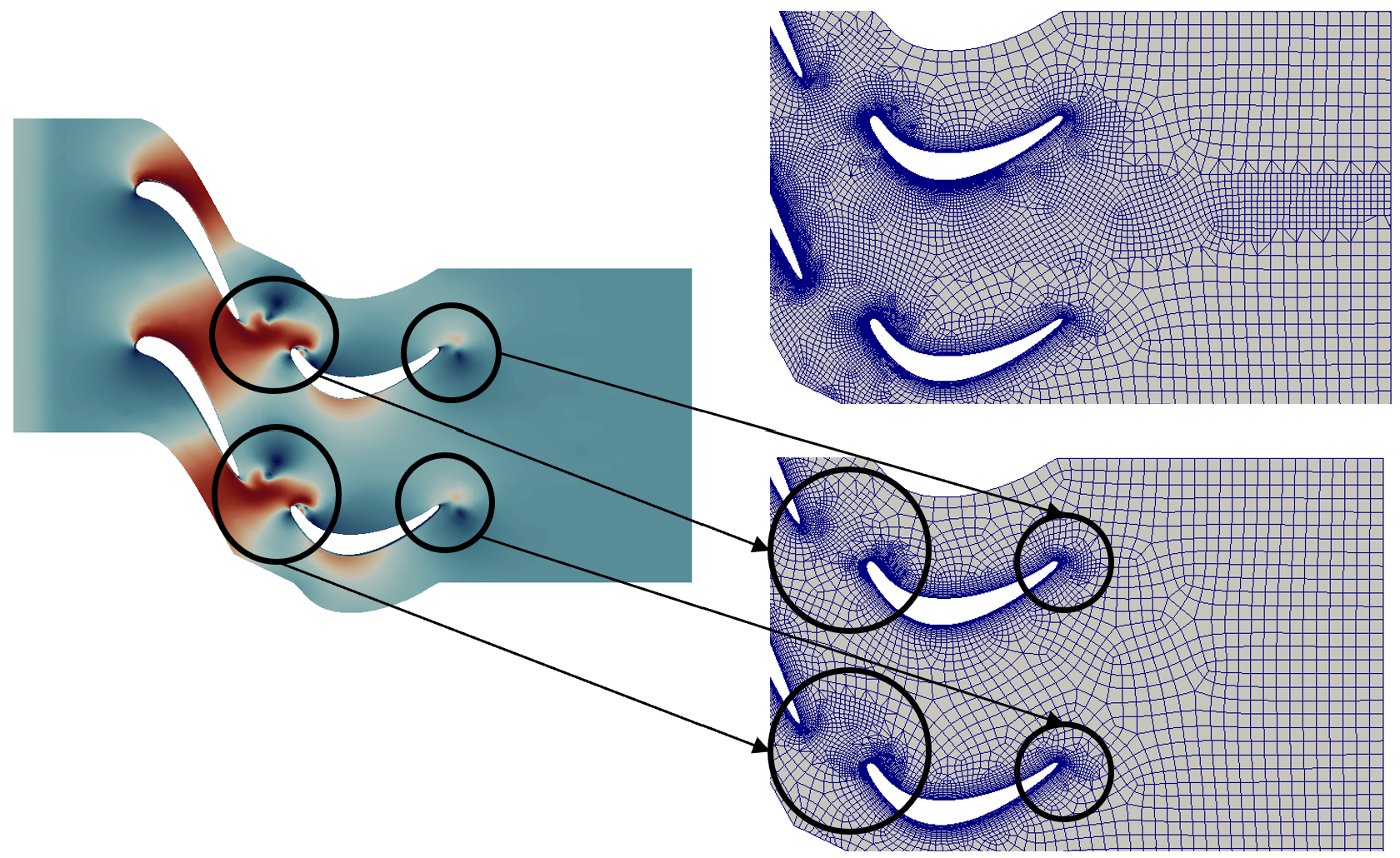}
	\caption{Flowfield for turbine blade domain (Left). Predictions for refinement made by ANN with four inputs as mesh cell center coordinates and non-dimensional velocity gradient (Top-Right) vs. ANN with four inputs as four non-dimensional quantities described in section \ref{sec:criteria} (Bottom-Right)}
	\label{fig:oldNN_newNN}
\end{figure}

We re-trained the same neural network architecture with a new training dataset consisting of 2.6M new input parameters and training labels evenly sampled from different time snapshots of all 3 problems shown in Figure \ref{fig:allproblems}. Re-training the network showed improved accuracy of 94\%. This improvement indicates that using proper non-dimensional quantities derived from flow is critical for the classifier's generalization. Figure \ref{fig:oldNN_newNN} shows prediction made using the old and new neural network models on turbine blade case. It is evident that the neural network trained on ensemble simulation data performs much better and targets specific regions for refinement, leading to improved computational efficiency.

\subsubsection{Boundary Detection}
\begin{figure}[htb]
	\centering
	\includegraphics[width=0.5\linewidth]{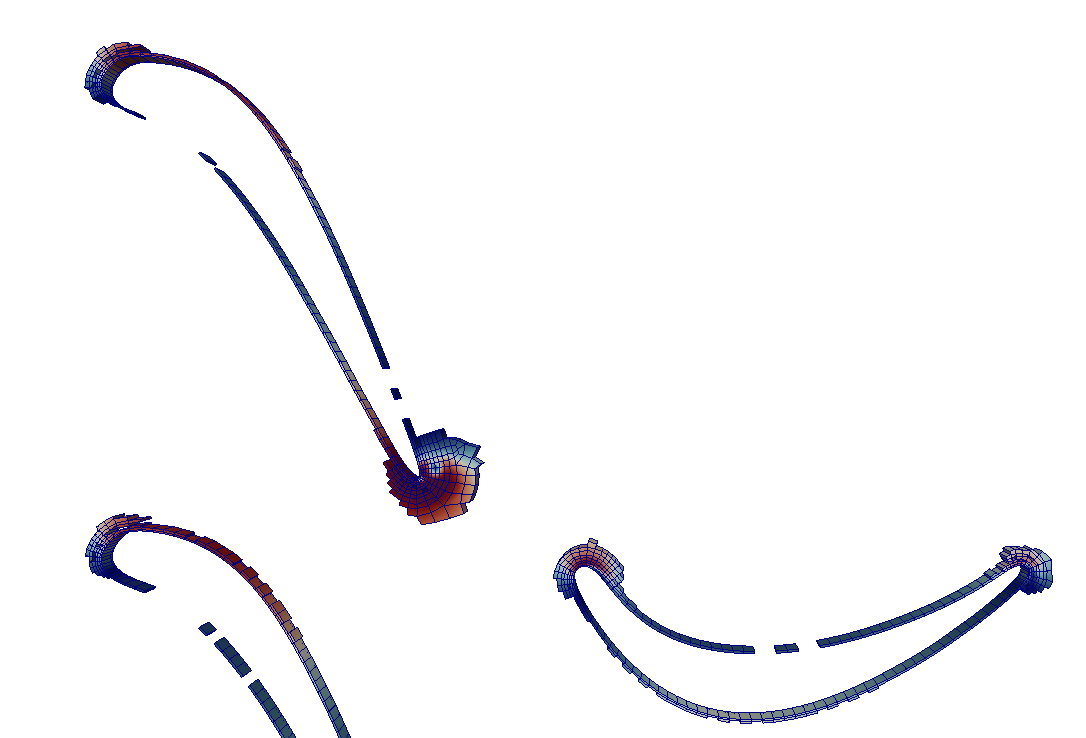}
	\caption{Boundaries predicted by trained convolutional neural network}
	\label{fig:CNN_Bndry}
\end{figure}

Smart AMR can be used to automate complex mesh optimization for CFD analyses, e.g., to predict the flowfield problem's boundaries.  In this study, we use a CNN classifier shown in Figure \ref{fig:CNN_Arch} to label boundary layer cells where turbulent structures exist. We used the same four non-dimensional quantities described above as different channels for the input layer. We trained this network on ensemble data generated from simulation problems shown in Figure \ref{fig:allproblems}. This dataset contains 2.6M training inputs and training labels. This network achieved a training accuracy of 92.5\% for predicting three different classes: i) internal cell refinement, ii) boundary layer cell refinement and iii) coarsening. Figure \ref{fig:CNN_Bndry} shows the trained classifier's prediction for turbine blade mesh. The classifier was able to pick boundaries of airfoils and nearest boundary layer cells, which are candidates for refinement. 

\section{Conclusion}
In this work, we introduced the smart AMR methodology to tackle adaptive refinement without optimal refinement criteria. We first discussed the algorithm used for the method. We introduced our object-oriented implementation. The application of proper non-dimensional feature extraction criteria suitable for different classes of problems and training operations used were discussed in detail. The feasibility of the method in performing adaptive mesh refinement was demonstrated on a few representative problems. Our preliminary studies indicated that smart refinement with a proper set of input and training is generalizable to similar nature problems. Finally, we presented an example of the application of smart AMR for flowfield boundary mesh optimization. Future works may include an extension to physical phenomena other than flow problems and extension to three-dimensional cases.

\section*{Acknowledgments}
This work has been supported by the U.S. Department of Energy Office of Sciences SBIR grant No. DE-SC0018077.

\bibliography{paper_scitech_2020}

\end{document}